\documentclass[twocolumn,showpacs,prb]{revtex4-1}
\usepackage{amsmath}
\usepackage{amssymb}
\usepackage{amsbsy}
\usepackage{xcolor}
\usepackage{soul}
\usepackage{graphicx}

\newcommand{\spvec}[1]{\ensuremath{\mathbf{#1}}}
\newcommand{\unitvec}[1]{\ensuremath{\mathbf{\hat{#1}}}}
\newcommand{\sptensor}[1]{\ensuremath{\boldsymbol{\mathbf{#1}}}}

\newcommand{\colvec}[1]{\ensuremath{\mathrm{#1}}}

\newcommand{\commentout}[1]{{}}
\newcommand{\half}{\hbox{$1\over2$}}

\newcommand{\rv}{\spvec{r}}
\newcommand{\Ev}{\spvec{E}}

\newcommand{\Bv}{\spvec{B}}
\newcommand{\Hv}{\spvec{H}}
\newcommand{\Pv}{\spvec{P}}
\newcommand{\Mv}{\spvec{M}}

\newcommand{\kv}{\spvec{k}}

\newcommand{\beq}{\begin{equation}}
\newcommand{\eeq}{\end{equation}}
\newcommand{\sji}{i}
\newcommand{\sje}{e}
\newcommand{\sjd}{\mathrm{d}}
\newcommand{\sjrme}{\mathrm{e}}

\newcommand{\rmm}{\textrm{m}}
\newcommand{\rmE}{\mathrm{E}}
\newcommand{\rmM}{\mathrm{M}}

\newcommand{\rmS}{\mathrm{S}}
\newcommand{\rmr}{\mathrm{r}}
\newcommand{\rml}{\mathrm{l}}
\newcommand{\CASR}{\mathcal{C}^{(\mathrm{ASR})}}

\begin{document}

\title{Resonance linewidth and inhomogeneous broadening in a metamaterial array}
\author{S. D. Jenkins}
\affiliation{School of Mathematics and Centre for Photonic
  Metamaterials, University of Southampton,
  Southampton SO17 1BJ, United Kingdom}
\author{J. Ruostekoski}
\affiliation{School of Mathematics and Centre for Photonic
  Metamaterials, University of Southampton,
  Southampton SO17 1BJ, United Kingdom}

\begin{abstract}
  We examine the effect of inhomogeneous broadening on the collective response of a planar metamaterial
  consisting of asymmetric split ring resonators. We show that such a response
  leads to a transmission resonance that can persist when the
  broadening of individual meta-atom resonance frequencies is roughly
  one half the frequency characterizing the split ring asymmetry. We also
  find that larger degrees of inhomogeneous broadening can drastically alter the
  cooperative response, destroying this resonance. The reduced effect
  of cooperative
  response due to inhomogeneous broadening may find applications
  in producing metamaterial samples that more closely mimic
  homogeneous magneto-dielectric
  medium with well-defined susceptibility and permittivity.
\end{abstract}

\date{\today}

\maketitle

\section{Introduction}
\label{sec:introduction}

There exists increasing experimental evidence that metamaterial
samples, consisting of arrays of sub-wavelength scale nano-structured
circuit elements, can be prepared in the limit where collective
interactions between the resonators play a vital role in determining
their electromagnetic (EM) responses.
For example, experiments on two-dimensional (2D) arrays of closely
spaced asymmetric split ring (ASR) meta-molecules have indicated the
presence of a high-quality transmission resonance, with a dramatic
sensitivity of the resonance linewidth to the number of ASRs in the
sample. \cite{FedotovEtAlPRL2010}
The transmission resonance was also observed to
  depend strongly on the spatial distribution of the mutually
  interacting resonators.
Where a sufficiently large, regular array of ASRs exhibits a high quality
transmission resonance,\cite{FedotovEtAlPRL2010}
introducing disorder in the elements' positions destroys the observed
spectral transmission window,\cite{papasimakis2009,SavoEtAlPRB2012}
further indicating the role collective EM interactions can play in
metamaterial dynamics.
We recently showed \cite{JenkinsLongPRB,JenkinsLineWidthArxiv} that
closely spaced ASR metamolecules interacting via a resonant EM field
exhibit collective eigenmodes with strongly suppressed resonance
linewidths.
The cooperative response yields the characteristic feature in the
experimentally observed enhanced quality factor of the transmission
resonance in Ref.~\onlinecite{FedotovEtAlPRL2010}.
Numerically analyzing the properties of collective modes with narrow
radiative resonance linewidths provided a physical explanation of this
phenomenon with an excellent agreement between the simulations and the
measurements.

In our previous study\cite{JenkinsLineWidthArxiv} showing how
  the transmission resonance observed by Fedotov \textit{et
    al}\cite{FedotovEtAlPRL2010} depends on the linewidth of a
  particular collective mode, all unit-cell
resonators were assumed to respond to EM fields identically.
In the preparation of metamaterial samples, fabrication defects,
however, may in general lead to variation in the geometry of
individual resonators.
The current oscillations supported by the unit-cell resonators would
therefore possess slightly different resonance frequencies, resulting
in inhomogeneously broadened metamaterial arrays.
Inhomogeneous broadening changes the conditions of the resonant
interaction processes.
This may impair the collective, coherent phenomena which are
potentially important in several applications and physical effects of
metamaterials such as lasing \cite{ZheludevEtAlNatPhot2008} and
providing precise control and manipulation of EM fields on a
sub-wavelength scale, as theoretically proposed in Ref.~\onlinecite{KAO10}
and experimentally observed in Ref.~\onlinecite{KaoEtAlNL2012} (for other related
studies, see for example Ref.~\onlinecite{SentenacPRL2008}).

In this work, we study how uncontrolled inhomogeneous
broadening of plasmonic resonators limits the observation of
collective phenomena in metamaterials.
We will examine under what conditions these collective effects can
still persist and potential implications of inhomogeneous broadening
on metamaterial applications.
In particular, we consider a rectangular 2D array of ASR
meta-molecules whose resonance frequencies are stochastically varied.
We numerically evaluate ensemble averages of the EM response of the
metamaterial over the stochastic distributions of the resonator
properties.
These show how increased inhomogeneous broadening inhibits the
response of the coherent collective modes responsible for the
transmission resonance observed in
Ref.~\onlinecite{FedotovEtAlPRL2010}.
Moreover, we find the effects of interactions between different
discrete resonator elements, which result in the narrowing of the
radiative resonance linewidth, are diminished as a function of an
increasing inhomogeneous broadening.
Our results therefore illustrate how maintaining uniformity in the
fabrication process is essential in designing new metamaterial based
devices whose applications rely on strong interactions between the
resonator elements and on a cooperative response.

The multiple scale spatial structure associated with nano-fabricated
resonators in metamaterial arrays, along with the wave nature of
scattered EM fields, poses a theoretical challenge for studies of
the response of these systems to resonant EM fields.
Interactions resulting from recurrent scattering events, in which a
field is scattered more than once by the same resonator, frequently
play a crucial role in the cooperative system responses.
\cite{Ishimaru1978,vantiggelen90,MoriceEtAlPRA1995,
  RuostekoskiJavanainenPRA1997L, RuostekoskiJavanainenPRA1997,
  JavanainenEtAlPRA1999,devries98,fermiline,muller01,pinheiro04,JenkinsLongPRB,JenkinsLineWidthArxiv,optlattice,dalibardexp}
While in an infinite, regular lattice, the translational symmetry can
be exploited to calculate approximate local field corrections in a
medium of discrete scatterers \cite{Kastel07}, recurrent
scattering processes are generally more difficult to model in
finite-sized systems with complex geometries.

However, since inhomogeneous broadening reduces the cooperative
effects arising from recurrent scattering, engineering a controlled
amount of inhomogeneous broadening into the metamaterial, e.g. via
geometrical variation of the resonators, may provide a practical means
to produce samples that are easier to design and theoretically analyze.
In particular, we find that with an increasing inhomogeneous
broadening the response of the system approaches that of standard
continuous medium electrodynamics.
Reducing cooperative effects is potentially important because several metamaterial
applications, such as diffraction-free lenses formed from a medium
with a negative refractive
index,\cite{SmithEtAlPRL2000,ShelbySci2001,SmithEtAlSCI2004} are
simplest to realize with a well-defined electric susceptibility and
magnetic permeability that, in many systems with complex geometries,
are only approximately achieved.\cite{szabo}

The effects of inhomogeneous broadening have previously been examined,
e.g., by Gorkunov \textit{et al} on the bulk properties of left-handed
materials in periodic infinite lattices,\cite{PhysRevE.73.056605} and
have been experimentally observed by Gollub \textit{et al}.\cite{Gollub}
In our study of ASR resonators we evaluate collective modes of a
finite lattice in order to investigate the effects of the
inhomogeneous broadening on the experimentally observed sample size
dependent transmission resonant linewidth narrowing.
\cite{FedotovEtAlPRL2010}

In our analysis we employ a general theoretical formalism of collective
interactions between a discrete set of plasmonic resonators, or
meta-atoms, mediated by the EM field that we developed in
Ref.~\onlinecite{JenkinsLongPRB}.
In the model, we assume each meta-atom exhibits a single mode of
current oscillation
that possesses appropriate electric and magnetic dipole moments.
Each meta-atom responds to EM fields exhibiting a specific resonance
frequency and coupling strength that are determined by its
characteristic design.
Starting from the Lagrangian and Hamiltonian formalism describing the
interaction of the EM field with polarization and
magnetization densities created by a charge distribution, we then
derived the coupled dynamics of the EM fields and the meta-atom dynamic
variables.\cite{JenkinsLongPRB} In a collection of meta-atoms, interactions with the EM field
mediate a dynamic coupling between
the meta-atoms and determine the collective dynamics within the
ensemble, resulting in distinct collective modes with corresponding
resonance frequencies and linewidths.
The analysis of collective response in terms of discrete resonators
also points to the direction of an interesting analogy between
resonators and a system of a cloud of atoms.
In atoms the electron transitions driven by an EM field create an
electric dipole moment, while in the case of circuit elements the
oscillating current generates both the electric and magnetic dipole
moments.
The model of Ref.~\onlinecite{JenkinsLongPRB} has previously been
successful in providing an excellent agreement between the theory and
experimental observations of cooperative transmission resonance
linewidth narrowing of ASR metamolecules.\cite{JenkinsLineWidthArxiv}

The remainder of this article is organized as follows.
We summarize the theoretical formalism we employ to describe
collective interactions within the metamaterial \cite{JenkinsLongPRB}
in Sec.~\ref{sec:model-coll-resp}.
In Sec.~\ref{sec:asymm-split-rings}, we describe the fundamental
building block of our metamaterical, the ASR, in the context of this
formalism.
The main results of the article, describing the effects of
inhomogeneous broadening on the collective response of the
metamaterial are presented in Sec.~\ref{sec:inhom-broad-}, and
conclusions follow in Sec.~\ref{sec:conclusion}.

\section{A Model for the Collective Response in Metamaterials}
\label{sec:model-coll-resp}

In order to incorporate the effects of strongly heterogeneous
metamaterial we describe the sub-wavelength structures of the medium
as discrete scatterers.\cite{JenkinsLongPRB}
Each unit-cell element, a meta-molecule, may also consist of sub-elements,
which we call meta-atoms.
While the general formalism of Ref.~\onlinecite{JenkinsLongPRB} allows
for multipole-field
radiation of the resonator unit elements, as a first approximation
here we consider each sub-wavelength-sized meta-atom simply as a
radiating dipole and ignore its multipole-field contribution.
Following our treatment in Ref.~\onlinecite{JenkinsLongPRB}, we assume
that each
meta-atom $j$, with its position vector defined by $\spvec{r}_j$,
supports a single eigenmode of current oscillation.
The dynamics of this current oscillation are determined by the dynamic
variable
$Q_j(t)$ with units of charge.
Each meta-atom exhibits an electric and magnetic dipole moment.
These may be expressed as
\begin{subequations}
  \label{eq:eDipDef}
  \begin{eqnarray}
    \spvec{d}_j &=& Q_j h_j \unitvec{d}_j \,\textrm{,} \\
    \spvec{m}_j &=& I_j A_j\unitvec{m}_j \,\textrm{,}
  \end{eqnarray}
\end{subequations}
respectively.
Here $I_j(t) = \sjd Q_j/ \sjd t$ denotes the current, and the directions of the
dipole moments are specified by the unit vectors $\unitvec{d}_j$ and
$\unitvec{m}_j$ with proportionality coefficients
$h_j$ and $A_j$ (with units of length and area, respectively) that
depend on the specific geometry of the resonators.
We assume the meta-atoms are designed such that the electric
quadrupole and higher order multipole contributions to the meta-atom
dynamics can be ignored.
Although each meta-atom possesses only electric and magnetic dipoles,
a meta-molecule of two or more meta-atoms in our model would exhibit a
non-vanishing quadrupole field.
While, in general, this quadrupole contribution is inaccurately
represented in the dipole approximation, in the case of the ASR
meta-molecules considered in
Refs.~\onlinecite{FedotovEtAlPRL2010,FedotovEtAlPRL2007} and in the
present study, the generated quadrupole field is notably suppressed
when compared to the corresponding dipolar field.
This has been indicated by
finite element simulations of Maxwell's equations within a single
meta-molecule.\cite{PapasimakisComm}
Additionally, the fact that the size of the meta-atoms is often
comparable to the spacing between them could result in a
correction to the coupling strength between neighboring elements
obtained in the point dipole approximation.
Nonetheless, this model in the dipole approximation was employed in
Ref.~\onlinecite{JenkinsLineWidthArxiv} to characterize the
cooperative linewidth narrowing responsible for the enhancement of
quality factor with system size observed in
Ref.~\onlinecite{FedotovEtAlPRL2010}, yielding excellent agreement with
experimental results.
In this section, we describe the key features of our theoretical
formalism that are required to describe the collective response of an
inhomogeneously broadened sample of ASRs to the EM field.
Details of the derivation are presented in
Ref.~\onlinecite{JenkinsLongPRB}.

We write the polarization and magnetization densities as a sum of
their contributions from the individual meta-atoms
\begin{subequations}
    \label{eq:totalPolMagDens}
  \begin{eqnarray}
    \Pv(\rv) &=&\sum_j \Pv_j(\rv) \,\textrm{,} \\
    \Mv(\rv) &=&\sum_j \Mv_j(\rv) \, \textrm{,}
  \end{eqnarray}
\end{subequations}
where the polarization and the magnetization of the resonator $j$ in the dipole approximation read
\begin{subequations}
  \begin{eqnarray}
    \spvec{P}_j(\rv,t) &\approx& \spvec{d}_j \delta(\rv-\rv_j) \,\textrm{,} \\
    \spvec{M}_j(\rv,t) &\approx& \spvec{m}_j \delta(\rv-\rv_j) \, \textrm{,}
  \end{eqnarray}
\end{subequations}
respectively.

An external beam with electric field
$\spvec{E}_{\mathrm{in}}(\spvec{r},t)$ and magnetic field
$\spvec{H}_{\mathrm{in}}(\spvec{r},t)$ with frequency
$\Omega_0$
impinges on the ensemble of meta-atoms.
The incident EM field drives the meta-atoms, generating dipole radiation
from the oscillating electric and magnetic dipoles.
The total radiation from the metamaterial array is the sum
of the scattered electric and magnetic fields from all the meta-atoms
\begin{subequations}
  \label{eq:TotalScatteredFields}
  \begin{eqnarray}
    \Ev_{\mathrm{S}}(\rv,t)=\sum_j \Ev_{\mathrm{S},j}(\rv,t) \,\textrm{,} \\
    \Hv_{\mathrm{S}}(\rv,t)=\sum_j \Hv_{\mathrm{S},j}(\rv,t) \, \textrm{,}
  \end{eqnarray}
\end{subequations}
where $\Ev_{\mathrm{S},j}(\rv,t)$ and $\Hv_{\mathrm{S},j}(\rv,t)$
denote the electric and magnetic field emitted by the meta-atom $j$.
The Fourier components of the scattered fields have the familiar
expressions of electric and magnetic dipole radiation,\cite{Jackson}
\begin{eqnarray}
  \spvec{E}^+_{\mathrm{S},j}(\spvec{r},\Omega) &=&
  \frac{k^3}{4\pi\epsilon_0} \int \sjd^3 r' \,
  \Bigg[\sptensor{G}(\spvec{r} - \spvec{r}',\Omega) \cdot
    \spvec{P}^+_{j}(\spvec{r}',\Omega)    \nonumber\\
    &&\qquad+ \frac{1}{c}
    \sptensor{G}_\times(\spvec{r}-\spvec{r}',\Omega) \cdot
    \spvec{M}^+_j(\spvec{r}',\Omega) \Bigg] , \label{efield}\\
     \spvec{H}^+_{\mathrm{S},j}(\spvec{r},\Omega) &=&
  \frac{k^3}{4\pi} \int \sjd^3 r' \,
  \Big[\sptensor{G}(\spvec{r} - \spvec{r}',\Omega) \cdot
  \spvec{M}^+_j(\spvec{r}',\Omega) \nonumber\\
  &&\qquad- c \sptensor{G}_\times(\spvec{r}-\spvec{r}',\Omega) \cdot
  \spvec{P}^+_j(\spvec{r}',\Omega) \Big]\,,\label{mfield}
\end{eqnarray}
where we have defined the positive and negative frequency components
of a real time varying quantity $V(t)$ such that for a Fourier component of
frequency $\Omega$, $V^{\pm}(\Omega) \equiv
\Theta(\pm\Omega)
V(\Omega)$, and hence $V(t) = V^+(t) +V^{-}(t)$ with $V^{-}(t) =
[V^{+}(t)]^*$.
Here $\sptensor{G}$ denotes the radiation kernel representing the
electric (magnetic) field emitted from an electric (magnetic) dipole.
\cite{Jackson}
The explicit expression for the corresponding radiated field from a
dipole $\spvec{v}$ reads
\begin{align}
  \sptensor{G}(\spvec{r},\Omega) &\cdot\spvec{v} =
  (\unitvec{r}\!\times\!\spvec{v}
  )\!\times\!\unitvec{r}
  \frac{\sje^{\sji kr}}{kr}+[3\unitvec{r}(\unitvec{r}\cdot\spvec{v})
  -\spvec{v}]\nonumber\\
  &\times \left[ \frac{1}{(kr)^3} - \frac{\sji
    }{(kr)^2}\right]\sje^{\sji kr}
  -{4\pi\over3}\delta(k\spvec{r}) \spvec{v}\,,
\label{eq:Green'sfunc}
\end{align}
where $\unitvec{r} \equiv \rv/r$ and $k \equiv \Omega/c$.
Similarly, $\sptensor{G}_{\times}(\spvec{r},\Omega)$ represents the
radiation kernel for the magnetic (electric) field of an electric
(magnetic) dipole source.\cite{Jackson}
Specifically, the corresponding radiated field from a dipole
$\spvec{v}$ yields
\begin{equation}
  \label{eq:CrossGreen}
  \sptensor{G}_{\times} (\spvec{r},\Omega)\cdot \spvec{v} =
  \frac{\sje^{\sji kr}}{kr} \left(1-\frac{1} {\sji kr}\right) \,\unitvec{r}
  \times \spvec{v}\,\textrm{.}
\end{equation}


The polarization and magnetization densities appearing in
Eqs.~\eqref{efield} and \eqref{mfield} are themselves driven by the
incident and scattered EM fields.
This driving, combined with the scattered EM fields, yield a coupled
set of equations for the resonators and EM fields that we derived in
Ref.~\onlinecite{JenkinsLongPRB}.
Current excitations in each meta-atom $j$ produces a field that
interacts with the current and charge oscillations that generated it.
Due to these self-generated fields, a meta-atom $j$ exhibits behavior
similar to that of an LC circuit with resonance frequency\cite{JenkinsLongPRB}
\begin{equation}
  \label{eq:resFreqDef}
  \omega_j \equiv \frac{1}{\sqrt{L_j C_j}} \,\textrm{,}
\end{equation}
where $C_j$ is an effective self-capacitance and the effective
self-inductance is $L_j$.
In this work, we consider an inhomogeneously broadened sample of $N$
ASRs.  Each ASR, $l$, ($l=1\ldots N$) consists of two meta-atoms
whose resonance frequencies $\omega_j$ ($j=2l-1,2l$) are centered
around the random frequency $\omega_0 + X_l$, where $X_l$ are
independent identically distributed random variables.
We assume the meta-atom resonance frequencies occupy a narrow band
about $\omega_0$, i.e. $|\omega_j-\omega_0|,X_l \ll \omega_0$.
The oscillating electric and magnetic dipoles
of an isolated meta-atom
radiate energy
at respective rates $\Gamma_{\rmE}$ and
$\Gamma_{\rmM}$,\cite{JenkinsLongPRB}
\begin{eqnarray}
  \Gamma_{\mathrm{E},j} &\equiv&
  \frac{h_j^{2}C_j\omega_j^{4}}{6\pi\epsilon_{0}c^3} \,\textrm{,}
  \label{eq:Gamma_EDef} \\
  \Gamma_{\mathrm{M},j} &\equiv& \frac{\mu_{0}A_j^{2}\omega_j^{4}}{6\pi c^3
    L_j} \label{eq:Gamma_MDef} \,\textrm{,}
\end{eqnarray}
resulting in the scattered fields $\spvec{E}_{\mathrm{S},j}$ and
$\spvec{H}_{\mathrm{S},j}$ [see Eqs.~\eqref{efield} and
\eqref{mfield}].
For simplicity, we assume that these radiative emission rates
$\Gamma_{\rmE}$ and $\Gamma_{\rmM}$ are independent of the resonator
$j$ and that they are dominated by the meta-atom resonance
frequencies, i.e. $\Gamma_{\rmE,j},\Gamma_{\rmM,j} \ll \Omega_0$.
We further assume the  resonance frequencies occupy a narrow bandwidth
around the central frequency of the incident field.

The dynamics of current excitations in the meta-atom $j$ may then be
described by $Q_j(t)$ [introduced in Eq.~\eqref{eq:eDipDef}] and
its conjugate momentum $\phi_j(t)$ (with units of magnetic flux
\cite{JenkinsLongPRB}).
In terms of the positive frequency components the equations of motion
read~\cite{JenkinsLongPRB}
\begin{eqnarray}
  \dot{Q}_j^{+} &=& \left( 1 - \sji \frac{\Gamma_{\rmM}}
    {\omega_j} \right)  \frac{\phi^{+}_j}{L_j}
  - \frac{\mu_0A_j}{L_j} \unitvec{m}_j \cdot \spvec{H}^+_{j,\mathrm{ext}}(\rv_j,t)
  \label{eq:saEqmQ} \\
  \dot{\phi}_j^{+} &=& -\left(1 - \sji
    \frac{\Gamma_{\rmE}}{\omega_j}\right) \frac{Q^{+}_j}{C_j} +
  h_j \unitvec{d}_j \cdot \spvec{E}^+_{j,\mathrm{ext}}(\rv_j,t) \, \textrm{,}
  \label{eq:saEqmphi}
\end{eqnarray}
where the fields generated externally to meta-atom $j$ that drive its
dynamics, $\spvec{E}^+_{j,\mathrm{ext}}(\rv,t)$ and
$\spvec{H}^+_{j,\mathrm{ext}}(\rv,t)$
are produced by the sums of the corresponding incident fields and the fields
scattered by all other meta-atoms in the metamaterial sample, $\sum_{j'\ne j}
  \spvec{E}^+_{\rmS,j'}(\rv,t)$ and $\sum_{j'\ne j}
  \spvec{H}^+_{\rmS,j'}(\rv,t)$, respectively.
The component of the external electric field
$\spvec{E}_{j,\mathrm{ext}}$ oriented along the dipole direction
$\unitvec{d}_j$ provides a net external electromotive force (EMF) $h_j
\unitvec{d}_j \cdot \spvec{E}_{j,\mathrm{ext}}^+(\rv_j,t)$
which drives $\phi_j(t)$.
Similarly, the component of the external magnetic field
$\spvec{H}_{j,\mathrm{ext}}$ along the magnetic dipole direction
$\unitvec{m}_j$ provides a net applied magnetic flux $A_j\mu_0 \unitvec{m}_j
\cdot \spvec{H}^+_{j,\mathrm{ext}}(\rv_j,t)$ that drives $Q_j(t)$.
In the absence of radiative emission and interactions with external
fields, current and charge oscillate within the meta-atom at the
resonance frequency $\omega_j$.
The meta-atom dynamics are therefore naturally described by the slowly
varying normal variables
\begin{equation}
  \label{eq:normalVariables}
  b_j(t) \equiv \frac{\sje^{\sji \Omega_0 t}}{\sqrt{2}}
  \left(\frac{Q_j(t)}{\sqrt{\omega_jC_j}} +\sji
    \frac{\phi_j(t)}{\sqrt{\omega_jL_j}}\right) \, \textrm{.}
\end{equation}
In the absence of
external field interactions and damping, $b_j$ oscillates with
frequency $(\omega_j - \Omega_0)$, i.e. $b_j(t) = b_j(0)
\exp\left[-i \left(\omega_j - \Omega_0\right) t\right]$.
For nonzero $\Gamma_{\mathrm{E}},\Gamma_{\mathrm{M}} \ll \Omega_0$,
losses and driving from the external field act to perturb
this oscillation.

The current oscillation dynamics in the meta-atom $j$, described by ${Q}_j(t)$ and $\phi_j(t)$
in Eqs.~\eqref{eq:saEqmQ} and~\eqref{eq:saEqmphi}, is driven by the incident field and the fields scattered from all the other meta-atoms and acts as a source of radiation that, in turn, drives the other meta-atoms.
The expressions for the scattered fields by polarization and magnetization densities [Eqs.~\eqref{efield} and~\eqref{mfield}] (generated by excitations in meta-atoms) and the expressions for the oscillating charge dynamics [Eqs.~\eqref{eq:saEqmQ} and~\eqref{eq:saEqmphi}] form a coupled set of equations, describing EM field mediated interactions between
the resonators. In terms of the normal variables $b_j$, these interactions may be represented by the set of equations,\cite{JenkinsLongPRB}
\begin{equation}
  \dot{\colvec{b}} = \mathcal{C} \colvec{b} + \colvec{f}_{\mathrm{in}} \textrm{ ,}
  \label{eq:rwa_b_eqm}
\end{equation}
where we have defined
\begin{equation}
  \label{eq:bColvec}
  \colvec{b}(t) \equiv \left(
    \begin{array}{c}
      b_1(t)\\
      b_2(t)\\
      \vdots\\
      b_{nN}(t)
    \end{array}
  \right) \, \textrm{,} \qquad
  \colvec{f}_{\mathrm{in}}(t) \equiv \left(
    \begin{array}{c}
      f_{1,\mathrm{in}} (t) \\
      f_{2,\mathrm{in}}(t) \\
      \vdots \\
      f_{nN,\mathrm{in}}(t)
    \end{array}
  \right) \,\textrm{.}
\end{equation}
The driving $f_{j,\mathrm{in}}$ of each meta-atom $j$ results
  from the EMF and magnetic flux induced by the incident
  fields.\cite{JenkinsLongPRB}
The component of the incident electric field
$\spvec{E}_{\mathrm{in}}$ parallel to the electric dipole orientation
$\unitvec{d}_j$ induces the EMF,
while the component of $\spvec{H}_{\mathrm{in}}$ along the magnetic
dipole orientation $\unitvec{m}_j$ provides an incident magnetic
flux.
Here we assume that
the meta-atom magnetic dipoles are aligned perpendicular to the incident
magnetic field, and thus only the EMF contributes to the driving of
each meta-atom, which is given by
\begin{equation}
  \label{eq:f_driving}
  \sje^{-\sji \Omega_0 t} f_{j,\mathrm{in}}(t) = \sji \frac{h_j}{\sqrt{2\omega_j L_j}} \unitvec{d}_j \cdot
  \spvec{E}_{\mathrm{in}}^+(\rv_j,t)
  \, \textrm{.}
\end{equation}
The coupling matrix between the meta-atoms in Eq.~\eqref{eq:rwa_b_eqm}
reads
\begin{equation}
  \mathcal{C}  =  -\sji \mathrm{\Delta} - \frac{\Gamma}{2} \mathrm{I}
  + \frac{1}{2} \left( \sji
    \mathcal{C}_{\mathrm{E}} +
    \sji \mathcal{C}_{\mathrm{M}}
    + \mathcal{C}_\times
      + \mathcal{C}_\times^{T} \right) \textrm{,}
  \label{eq:C_rwa}
\end{equation}
where $\mathrm{I}$ represents the identity matrix.
Here the detunings of the incident field from the meta-atom resonances
are contained in the diagonal matrix $\mathrm{\Delta}$ with elements
\begin{equation}
  \Delta_{j,j'} \equiv \delta_{j,j'} \left(\omega_j-\Omega_0\right)\,\textrm{,}
  \label{eq:DeltaMatDef}
\end{equation}
and the energy carried away from individual meta-atoms by the
scattered fields manifests itself in the decay rate
\begin{equation}
  \label{eq:GammaMatDef}
  \Gamma \equiv \Gamma_{\rmE} +
    \Gamma_{\rmM} + \Gamma_{\mathrm{O}}
\end{equation}
appearing
in the diagonal elements of $\mathcal{C}$.
We account for non-radiative, e.g. ohmic losses, by introducing
phenomenological decay rate $\Gamma_{\mathrm{O}}$.
The multiple scattering processes are included in the terms $\mathcal{C}_{\rmE}$, $\mathcal{C}_{\rmM}$,
and $\mathcal{C}_{\times}$, which generate interaction between the
meta-atom dynamic variables
The matrices $\mathcal{C}_{\rmE}$ and $\mathcal{C}_{\rmM}$
characterize the electric dipole-dipole and magnetic dipole-dipole
interactions, respectively.
Additionally, the interaction embodied by $\mathcal{C}_\times$
arises from the \emph{electric} field emitted by the \emph{magnetic}
dipole of one atom driving the electric dipoles of the others.
Similarly, $\mathcal{C}_{\times}^T$ results from the \emph{magnetic}
field produced by the meta-atoms' \emph{electric} dipoles impinging
on the magnetic dipoles of all the other meta-atoms.
Because the interaction matrices $\mathcal{C}_{\rmE |\rmM |\times}$
govern interactions between distinct meta-atoms, their diagonal
elements are zero.
Their off diagonal elements are given by
\cite{JenkinsLongPRB}
\begin{eqnarray}
  \label{eq:GreensElecMat}
  \left[\mathcal{C}_{\rmE}\right]_{j,j'} &=& \frac{3}{2}\Gamma_{\rmE}\,\unitvec{d}_j \cdot
  \sptensor{G}(\spvec{r}_j  - \spvec{r}_{j'},\Omega_0 )  \cdot
  \unitvec{d}_{j'},\\
  \label{eq:GreenMagMat}
  \left[\mathcal{C}_{\rmM}\right]_{j,j'}& =& \frac{3}{2} \Gamma_{\rmM} \, \unitvec{m}_j \cdot
  \sptensor{G}(\spvec{r}_j - \spvec{r}_{j'},\Omega_0 ) \cdot \unitvec{m}_{j'},\\
  \label{eq:emCrossMat}
  \left[\mathcal{C}_\times \right]_{j,j'} &=& \frac{3}{2}\bar{\Gamma}\, \unitvec{d}_j\cdot
  \sptensor{G}_\times(\spvec{r}_j -
  \spvec{r}_{j'},\Omega_0) \cdot
  \unitvec{m}_{j'}\,,
\end{eqnarray}
where $\bar{\Gamma} \equiv \sqrt{\Gamma_{\rmE} \Gamma_{\rmM}}$ is the
geometric mean of the electric and magnetic dipole emission rates.

\section{Asymmetric Split Ring Resonators}
\label{sec:asymm-split-rings}

In order to investigate the effects of inhomogeneous broadening
of meta-atom resonance frequencies on a metamaterial's collective EM
response, we consider an ensemble of asymmetric meta-molecules
arranged in a regular lattice.
To facilitate our description of this EM response, in this section we
summarize the behaviour of a single ASR in the context of the model
presented in Sec.~\ref{sec:model-coll-resp}.

An ASR is a variation on the split ring resonator used to produce bulk
metamaterials with negative indices of refraction.\cite{SmithEtAlPRL2000,ShelbySci2001,SmithEtAlSCI2004}
The meta-atoms of an ASR consist of two separate
concentric circular arcs labeled by $j \in\{\rml,\rmr\}$ and separated by
$\spvec{u} \equiv \spvec{r}_{\rmr} - \spvec{r}_{\rml}$.
The current oscillations in meta-atoms produce electric dipoles
with orientation $\unitvec{d}_{\rmr} = \unitvec{d}_{\rml} = \unitvec{d}$
associated with charge oscillating between the ends of the arcs.
Owing to the curvature of the meta-atoms, these currents also produce
magnetic dipoles with opposite orientations $\unitvec{m}_{\rmr} =
-\unitvec{m}_{\rml} = \unitvec{m}$ where $\unitvec{d}\perp\spvec{u}$ and
$\unitvec{m} \perp\spvec{u},\unitvec{d}$.
An asymmetry between the rings, e.g., resulting from a difference in
arc length, manifests itself as a difference in resonance
frequencies with
\begin{eqnarray}
  \label{eq:4}
  \omega_{\rmr} = \omega_0 + \delta\omega \\
  \omega_{\rml} = \omega_0 - \delta\omega
\end{eqnarray}

To analyze the dynamics of a single ASR unit-cell resonator consisting
of two meta-atoms, we apply the formalism presented in
Sec.~\ref{sec:model-coll-resp}.
According to (\ref{eq:rwa_b_eqm}), the normal variables $b_{\rmr}$ and
$b_{\rml}$  that describe the current oscillations in the right and
left meta-atoms, respectively, are coupled by the EM fields so that they
evolve according to
\begin{equation}
  \label{eq:SSRDynEq}
  \left(
    \begin{array}{c}
      \dot{b}_r \\
      \dot{b}_l
    \end{array}
  \right) = \CASR
  \left(
    \begin{array}{c}
      b_r \\
      b_l
    \end{array}
  \right)
  +
  \left(
    \begin{array}{c}
      f_{r,\mathrm{in}} \\
      f_{l,\mathrm{in}}
    \end{array}
  \right)\,.
\end{equation}
Here $\CASR$ denotes the specific coupling matrix
between the two meta-atoms that depends of the radiative electric dipole --
electric dipole, magnetic dipole -- magnetic dipole, and electric
dipole -- magnetic dipole interaction processes between the two
meta-atoms [See Eqs.~\eqref{eq:C_rwa}-\eqref{eq:emCrossMat}.].
On the other hand, the incident field produces the driving terms
$f_{j,\mathrm{in}}$ for each meta-atom $j=\rml,\rmr$ [See
Eq.~\eqref{eq:f_driving}.].

To analyze the modes of the ASR, we consider the dynamics
of symmetric $c_+$ and antisymmetric $c_-$ modes of oscillation
defined by
\begin{equation}
  \label{eq:c_pm_def}
  c_\pm \equiv \frac{1}{\sqrt{2}} \left(b_{\rmr} \pm b_{\rml} \right) \textrm{.}
\end{equation}
The
oscillations $c_{\pm}$ represent the eigenmodes of the ASR
in the absence of asymmetry $\delta\omega=0$.
By diagonalizing $\CASR$ with $\delta\omega=0$, one finds the
eigenvalues of the modes $c_\pm$,
\begin{equation}
  \label{eq:CASR_eigenvalues}
  \lambda_\pm = -i\left(\omega_0-\Omega_0 \pm \Delta\right) -
  \frac{\gamma_\pm}{2} \,\textrm{.}
\end{equation}
The interaction between the elements shifts the two collective
  resonance frequencies by equal and opposite amounts $\Delta$ and
  results in the decay rates $\gamma_{\pm}$, where the coefficients $\Delta$ and $\gamma_\pm$ depend on the radiative interactions between the two meta-atoms.\cite{JenkinsLongPRB,JenkinsLineWidthArxiv}
When the spacing between the arcs $u \ll \lambda$ (~$\lambda = 2\pi
c/\Omega_0$~), the decay rates simplify to
\begin{subequations}
  \begin{eqnarray}
    \label{eq:gamma_p}
    \gamma_+ &=& 2\Gamma_{\mathrm{E}}+\Gamma_{\mathrm{O}} \textrm{ ,}
    \\
    \label{eq:gamma_m}
    \gamma_- &=& 2\Gamma_{\mathrm{M}} + \Gamma_ {\mathrm{O}} \textrm{ .}
  \end{eqnarray}
\end{subequations}
In this limit, the symmetric mode, possessing a net electric dipole,
emits electric dipole radiation, and the antisymmetric mode,
possessing a net magnetic dipole, emits magnetic dipole radiation.
We therefore may refer to symmetric and antisymmetric oscillations as
electric and magnetic dipole excitations, respectively.

A nonzero asymmetry, $\delta\omega \ne 0$, tends to couple the
symmetric and anti-symmetric oscillations.
One finds that, when driven by an external
field, these oscillations in a single ASR evolve
as\cite{JenkinsLongPRB,JenkinsLineWidthArxiv}
\begin{equation}
  \label{eq:cpmEqs}
  \dot{c}_{\pm} = \left[-\sji\left(\omega_0\pm\Delta - \Omega_0\right)
    - \frac{\gamma_\pm}{2}
  \right]c_\pm - \sji \delta\omega c_{\mp} + F_\pm \,\textrm{,}
\end{equation}
where the driving terms $F_\pm = (f_{\rmr,\mathrm{in}} \pm f_{\rml,\mathrm{in}})/\sqrt{2}$.
The symmetric and antisymmetric oscillations are driven
purely by the electric and magnetic fields, respectively, and when the
meta-atom separation $u\ll\lambda$,  $F_+
\propto \unitvec{d} \cdot \Ev_{\mathrm{in}}^+(\spvec{R},t)$ and $F_-
\propto \unitvec{m} \cdot \Bv_{\mathrm{in}}^+(\spvec{R},t)$, where
$\spvec{R}$ is the center of mass of the ASR.
Therefore, an incident field with
$\spvec{E}_{\rm in} \parallel \unitvec{d}$ and $\spvec{B}_{\rm in} \perp
\unitvec{m}$ only excites the symmetric mode when
$\delta\omega =0$.
However for $\delta\omega \ne 0$, the asymmetry couples the symmetric
and antisymmetric modes, and this incident field
can resonantly pump the anti-symmetric magnetic mode via an effective
two-photon transition.\cite{JenkinsLineWidthArxiv}

\begin{figure}
 \centering
 \includegraphics[width=0.95\columnwidth]{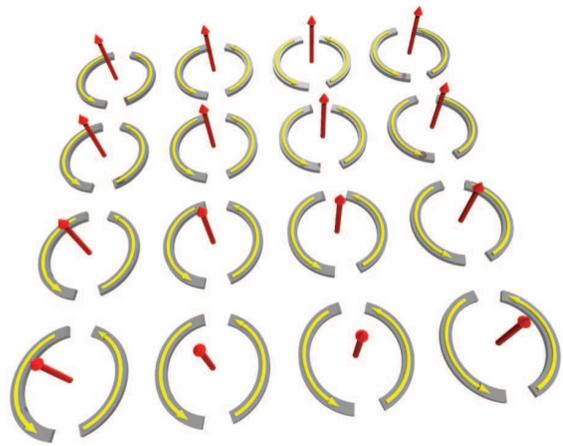}
 \caption{A schematic illustration of an array of ASR metamolecules
   excited in the uniform phase magnetic mode.
   The meta-atom currents in each ASR oscillate with opposite phases,
   producing magnetic dipoles represented by the red arrows.
 }
 \label{fig:mag_mode_schematic}
\end{figure}

In this article, we consider a 2D metamaterial comprised of ASRs
arranged in a regular array.
The fields scattered from each ASR then mediate interactions between
them, resulting in collective modes of oscillation distributed over
the array, each with its own resonance frequency and decay rate.
Figure~\ref{fig:mag_mode_schematic} provides a schematic illustration
of a mode consisting primarily of magnetic dipoles oscillating in
phase throughout the
metamaterial.\cite{JenkinsLongPRB,JenkinsLineWidthArxiv}
We showed in Ref.~\onlinecite{JenkinsLineWidthArxiv} that, for
sufficiently large array, such a mode radiates more slowly than the
magnetic excitation of a single ASR and is responsible for the transmission
resonance observed by Fedotov \textit{et al}.\cite{FedotovEtAlPRL2010}.
When an incident EM field whose magnetic field is perpendicular to the
ASR magnetic dipoles impinges on the array, this mode cannot be
excited directly. But, the presence of an asymmetry provides a
coupling between electric and magnetic dipoles allowing it to be
driven.
Driving of the metamaterial's collective modes are responsible for the
cooperative response that yields phenomena such as the
observed transmission resonance.\cite{FedotovEtAlPRL2010,JenkinsLineWidthArxiv}

\section{Inhomogeneous Broadening }
\label{sec:inhom-broad-}

In this section, we study the effects of inhomogeneous broadening on
the cooperative EM response of an array of ASRs.
Here inhomogeneous broadening refers to a statistical uncertainty in
the resonance frequencies in individual ASR meta-molecules.
Such an uncertainty may result, for example, from imperfections in the
manufacturing processes which yield meta-atoms whose shape varies
slightly from the design specifications.
Collective modes that are phase matched with an incident EM field have
been shown to be responsible for transmission resonances
\cite{JenkinsLineWidthArxiv} that have been observed
experimentally.\cite{FedotovEtAlPRL2007,FedotovEtAlPRL2010}
We will illustrate the response of a regular array of ASRs to an
incident plane wave, and show that broadening adversely affects the
characteristics of the response responsible for observed
resonances.

We consider an ensemble of ASR meta-molecules whose constituent
meta-atoms are separated by $\spvec{u}=u\unitvec{e}_x$ with electric
dipoles oriented along $\unitvec{d} = \unitvec{e}_y$ such that a
symmetric oscillation in a single ASR produces an electric dipole
along $\unitvec{d}$, and an antisymmetric oscillation produces a
magnetic dipole along $\unitvec{m}=\unitvec{e}_z$.
The ASRs are arranged in an $N_x \times N_y$ 2D square lattice with
lattice spacing $a$ and lattice vectors $\spvec{a}_1=a\unitvec{e}_x$
and $\spvec{a}_2 =a \unitvec{e}_y$.
The sample is illuminated by a cw plane wave $\Ev_{\rm in}^+(\rv)=
\half {\cal E} \unitvec{e}_y e^{i\kv \cdot \rv}$ with $\spvec{k} =k
\unitvec{e}_z$, coupling exclusively to the electric dipole moments of
the ASRs.

The vector of $2N_xN_y$ normal variables describing the state of
current oscillations in each meta-atom obeys the coupled equations of
motion in Eq.~\eqref{eq:rwa_b_eqm}, where the matrix $\mathcal{C}$
[See~Eq.~\eqref{eq:C_rwa}.] arises from the meta-atom interactions
mediated by the EM field.
The metamaterial therefore exhibits $2N_x N_y$ collective modes
of oscillation corresponding to the eigenvectors $\colvec{v}_i$
($i=1\ldots 2N_xN_y)$ of interaction matrix
$\mathcal{C}$.\cite{JenkinsLongPRB}
Each eigenmode $i$ possesses a particular resonance frequency
$\Omega_i$ and decay rate $\gamma_i$ given in terms of the eigenvalue
$\lambda_i$ by
\begin{subequations}
  \label{eq:eigenvals}
  \begin{eqnarray}
    \Omega_i &=& -\operatorname{Im}(\lambda_i) + \Omega_0\,\textrm{,} \\
    \gamma_i &=& -2\operatorname{Re}(\lambda_i) \,\textrm{,}
  \end{eqnarray}
\end{subequations}
respectively.

Since the incident field drives all ASRs uniformly, it couples most
effectively to the collective modes in which all of the metamolecules
oscillate in phase.
The two modes of particular interest are the uniform electric and
uniform magnetic modes.
In the absence of an asymmetry ($\delta\omega=0$), the incident field
drives the uniform electric mode which, owing to the electric dipole
orientations, emits strongly into the $\pm \unitvec{e}_z$ directions.
The electric dipoles oscillating in phase are responsible for
reflection from the metamaterial.
By contrast, the magnetic dipoles in the uniform magnetic mode,
illustrated in Fig.~\ref{fig:mag_mode_schematic}, emit
into the plane of the meta-material array, and for sufficiently large
lattices of closely spaced ASRs, the magnetic dipole radiation becomes
trapped.
This results in a suppressed radiative decay rate of the uniform
magnetic mode.\cite{JenkinsLineWidthArxiv}
Introduction of an asymmetry ($\delta\omega \ne 0$) provides an
effective coupling between these two collective modes similar to the
coupling between the symmetric and antisymmetric modes in a single
ASR.
The collective magnetic mode can thus be resonantly excited at the
expense of the electric dipoles resulting in a transmission resonance
\cite{JenkinsLineWidthArxiv,FedotovEtAlPRL2007} whose
quality factor increases with the size of the array, as observed by
Fedotov \textit{et al}. \cite{FedotovEtAlPRL2010}

\begin{figure}
  \centering
  \includegraphics[width=0.9\columnwidth]{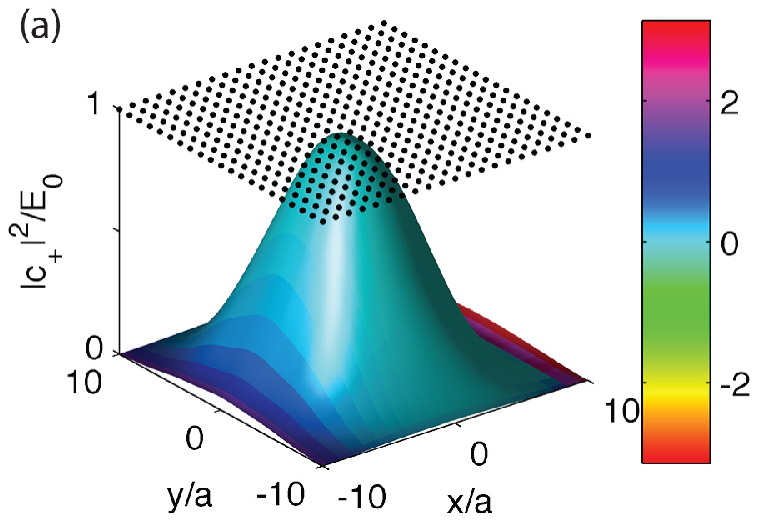}
  \includegraphics[width=0.9\columnwidth]{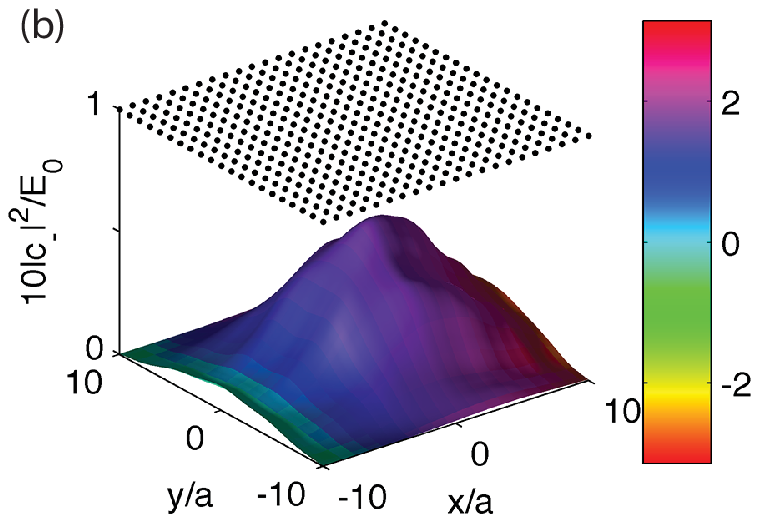}
  \caption{The uniform electric mode $\colvec{v}_{\sjrme}$ of a
    homogeneously  broadened  (~$\sigma=0$~)  $21 \times   21$ ASR
    square lattice.
    (a) The electric dipole excitations $|c_{+,l}|^2$ and (b) the
    magnetic dipole excitations $|c_{-,l}|^2$ of the uniform electric
    mode $\colvec{v}_{\sjrme}$.
    The phases of the electric ($c_{+,l}$) and magnetic ($c_{-,l}$)
    dipole excitations are indicated by the color of the surfaces in
    (a) and (b), respectively.
    The black dots indicate the positions of the ASRs in the array.
    The excitations  $|c_{\pm,l}|^2$ are normalized to the peak
    ASR excitation $E_0 = \max_l (|c_{+,l}|^2+|c_{-,l}|^2)$.
    The vertical scale of panel (b) was amplified by a factor of ten
    to render the magnetic dipole excitations $|c_{-,l}|^2$ visible.
    The vertices on the plots correspond to the ASR positions.
    The lattice spacing
    $a=0.28\lambda$, meta-atom separation within an ASR
    $u=0.12\lambda$, $\Gamma_{\mathrm{E}}=\Gamma_{\mathrm{M}}$, and
    asymmetry parameter $\delta\omega = 0.3\Gamma$.}
  \label{fig:uniformElectricMode}
\end{figure}
\begin{figure}
  \centering
  \includegraphics[width=0.9\columnwidth]{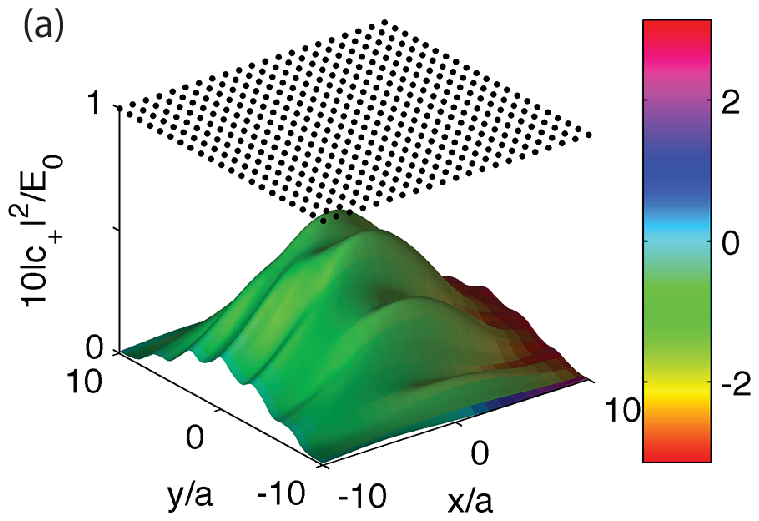}
  \includegraphics[width=0.9\columnwidth]{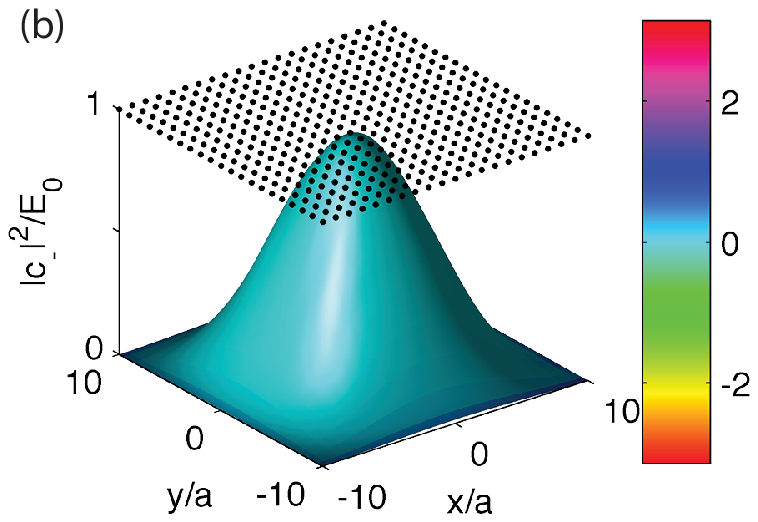}
  \caption{The uniform magnetic mode $\colvec{v}_{\rmm}$ of a homogeneously
    broadened  (~$\sigma=0$~)  $21 \times   21$ ASR square lattice.
    (a) The electric dipole excitations $|c_{+,l}|^2$ and (b) the
    magnetic dipole excitations $|c_{-,l}|^2$ of the uniform electric
    mode $\colvec{v}_{\sjrme}$.
    The phases of the electric ($c_{+,l}$) and magnetic ($c_{-,l}$)
    dipole excitations are indicated by the color of the surfaces in
    (a) and (b), respectively.
    The black dots indicate the positions of the ASRs in the array.
    The vertical scale of (a) was amplified by a factor of ten
    to render the electric dipole excitations $|c_{+,l}|^2$ visible.
    All parameters are as in Fig.~\ref{fig:uniformElectricMode}.
  }
  \label{fig:uniformMagneticMode}
\end{figure}

Specifically, the uniform magnetic mode is the eigenmode
$\colvec{v}_{\rmm}$
of $\mathcal{C}$ [See Eq.~\eqref{eq:C_rwa}.] which maximizes the
overlap $O_{\rmm}(\colvec{b}_A)$
with the pure magnetic excitation $\colvec{b}_{\mathrm{A}}$ in which
all meta-atoms are excited with equal amplitude and the two meta-atoms
in each split ring oscillate out of phase.
Similarly, the uniform electric mode is the eigenvector $\colvec{v}_{\sjrme}$ of
$\mathcal{C}$ that maximizes the overlap $O_{\sjrme}(\colvec{b}_{\rmS})$ with
the pure electric excitation $\colvec{b}_{\rmS}$ for which all current oscillations
oscillate in phase with equal amplitude.
Explicitly, these column vectors of $2N_xN_y$ elements are given by
\begin{equation}
  \label{eq:b_A_and_b_S_def}
  \colvec{b}_{\mathrm{A}} \equiv
  \left(
    \begin{array}{c}
      +1 \\
      -1 \\
      \vdots \\
      +1 \\
      -1
    \end{array}
  \right)\textrm{,}
  \qquad
  \colvec{b}_{\mathrm{S}} \equiv
  \left(
    \begin{array}{c}
      1 \\
      1 \\
      \vdots \\
      1 \\
      1
    \end{array}
  \right) \textrm{ .}
\end{equation}
The alternating signs of the elements of $b_{\mathrm{A}}$ indicate the
relative phase of the oscillations in each meat-atom of an ASR.
We define the overlap of mode $\colvec{v}_{\rmm/\sjrme}$ with an
arbitrary excitation $\colvec{b}$ as
\begin{equation}
  O_{\rmm/\sjrme}(\colvec{b}) \equiv \frac{ | \colvec{v}_{\rmm/\sjrme}^T \colvec{b} |^2}
  {\sum_i | \colvec{v}_i^T \colvec{b} |^2} \label{eq:overlapDef} \,\textrm{,}
\end{equation}
where the index $i$ is summed over all the eigenmodes of the
interaction matrix $\mathcal{C}$.
The uniform electric and uniform magnetic modes for a $21 \times 21$
array of ASRs with a nonzero asymmetry parameter
$\delta\omega=0.3\Gamma$ are shown in
Figs.~\ref{fig:uniformElectricMode} and
\ref{fig:uniformMagneticMode}, respectively.
We used the experimental value for the lattice spacing from
Ref.~\onlinecite{FedotovEtAlPRL2010} and the estimate the asymmetry
parameter $\delta\omega \simeq 0.3 \Gamma$ from the relative arc
lengths of the ASR meta-atoms studied by Fedotov \textit{et
  al}.\cite{FedotovEtAlPRL2007,FedotovEtAlPRL2010}
The ohmic loss rate $\Gamma_{\mathrm{O}}$ was fitted so that
the quality factor of the uniform magnetic mode as a function of
system size matched the experimental observations.
\cite{JenkinsLineWidthArxiv}
Where the state of the ensemble is characterized by
the vector of meta-atom normal variables $\colvec{b}$ [See Eq.~\eqref{eq:bColvec}.], the symmetric (electric) and
antisymmetric (magnetic) oscillations of an ASR $l$ (~$l=1,\ldots
N_xN_y$~) are represented by $c_{+,l}$ and $c_{-,l}$, respectively,
where
\begin{equation}
  \label{eq:2}
  c_{\pm,l} \equiv \frac{1}{\sqrt{2}}(b_{2l-1} \pm b_{2l}) \textrm{ .}
\end{equation}
The respective symmetric (electric dipole)  and antisymmetric (magnetic
dipole) excitation energies in ASR $l$ are proportional to
$|c_{+,l}|^2$ and $|c_{-,l}|^2$.
The asymmetry in the ASRs causes a mixing of the electric and
magnetic dipoles, producing a slight electric dipole excitation in the
uniform magnetic mode $\colvec{v}_{\rmm}$ and a slight magnetic
excitation of the
electric mode $\colvec{v}_{\sjrme}$.
In the example illustrated in Figs.~\ref{fig:uniformElectricMode} and
\ref{fig:uniformMagneticMode}, when the ohmic loss rate in each
meta-atom is $\Gamma_{\mathrm{O}} \simeq 0.14\Gamma$, the electric mode
has an enhanced decay
rate $\gamma_{\mathrm{e}} = 2.7 \Gamma$ and the magnetic mode has a
suppressed decay rate $\gamma_{\rmm} = 0.31\Gamma$ with respect
to the total isolated single meta-atom decay rate $\Gamma$.
Due to the suppressed decay rate of $\colvec{v}_{\rmm}$ and its small
electric dipole component, this mode can be resonantly excited by  the
incident field.

\begin{figure}
  \centering
  \includegraphics[width=0.9\columnwidth]{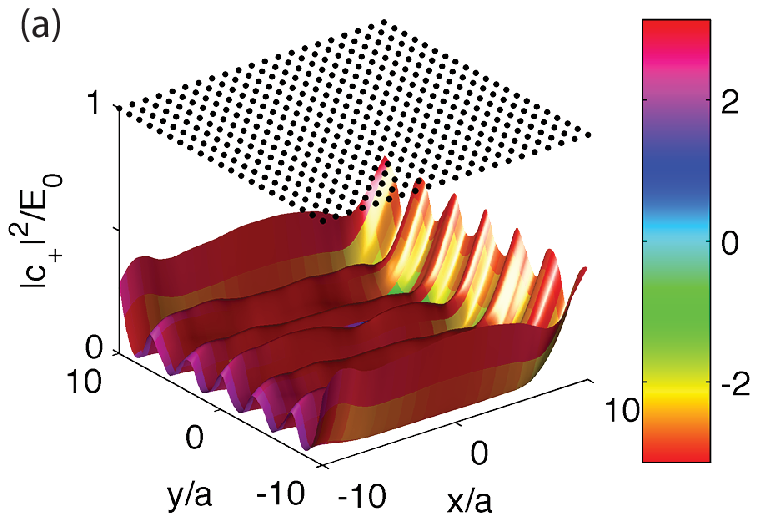}
  \includegraphics[width=0.9\columnwidth]{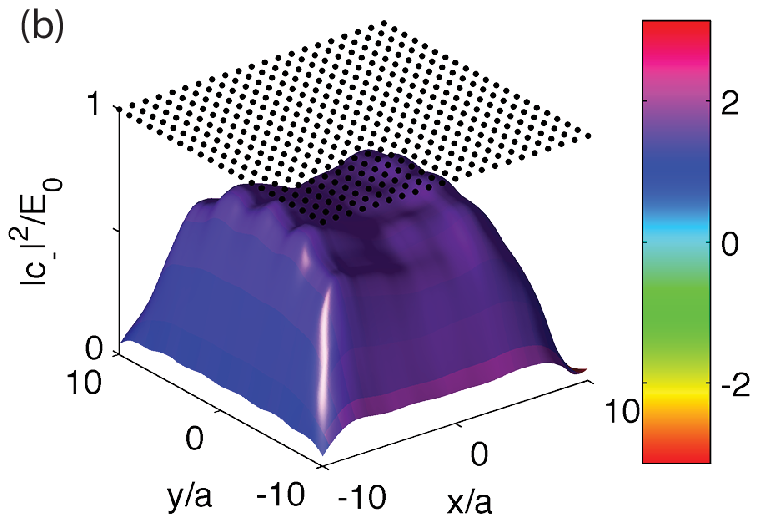}
  \caption{The response of a homogeneously broadened (~$\sigma=0$~)
    $21 \times 21$ ASR square lattice to an
    incident plane wave electric field resonant on the uniform
    collective magnetic mode $\colvec{v}_{\rmm}$, showing the excitation of
    a uniform magnetic dipole excitation at the expense of electric
    dipoles.
    (a) The electric dipole excitations $|c_{+,l}|^2$ and (b) the
    magnetic dipole excitations $|c_{-,l}|^2$ of the response.
    The phases of the electric ($c_{+,l}$) and magnetic ($c_{-,l}$)
    dipole excitations are indicated by the color of the surfaces in
    (a) and (b), respectively.
    The black dots indicate the positions of the ASRs in the array.
    The ohmic loss rate is $\Gamma_{\mathrm{O}} = 0.14\Gamma$.
    All other parameters are as in Fig.~\ref{fig:uniformElectricMode}.
  }
  \label{fig:response-nonBroadened}
\end{figure}

We illustrate this phenomenon in Fig.~\ref{fig:response-nonBroadened},
where we show the steady state response [See Eq.~\eqref{eq:rwa_b_eqm}.],
\begin{equation}
  \label{eq:b_r_def}
  \colvec{b}_{\mathrm{r}} \equiv -
  \mathcal{C}^{-1}\colvec{f}_{\mathrm{in}} \,\textrm{,}
\end{equation}
of an array whose resonance frequencies are
not inhomogeneously broadened.
The right and left meta-atoms
of ASR $l$ in such an array have respective resonance frequencies
$\omega_{2l-1} = \omega_{0} + \delta\omega$ and $\omega_{2l} =
\omega_0 - \delta\omega$ centered around $\omega_0$.
The driving field is resonant on the uniform magnetic mode, and the
asymmetry in the split rings $\delta\omega = 0.3 \Gamma$ facilitates
the phase-coherent excitation of the magnetic dipoles at the expense
of the electric dipoles.
Panels (a) and (c) of Fig.~\ref{fig:response-nonBroadened} illustrate
that the magnetic dipoles are much more strongly excited than the
electric dipoles in the bulk of the array, and that these magnetic
dipoles oscillate in phase.
The more excited of the weak electric dipoles also oscillate in phase,
thus facilitating the driving of this excitation by the uniform
incident field.
Although other collective modes of the system are excited, more than
$60\%$ of the excitation energy resides in the uniform magnetic mode
$\colvec{v}_{\rmm}$.
In the absence of ohmic losses, one can optimize the asymmetry
parameter $\delta\omega$ in large lattices so that over $98\%$ of the
excitation energy resides in the uniform magnetic mode.
\cite{JenkinsLineWidthArxiv}

\begin{figure}
  \centering
  \includegraphics[width=0.9\columnwidth]{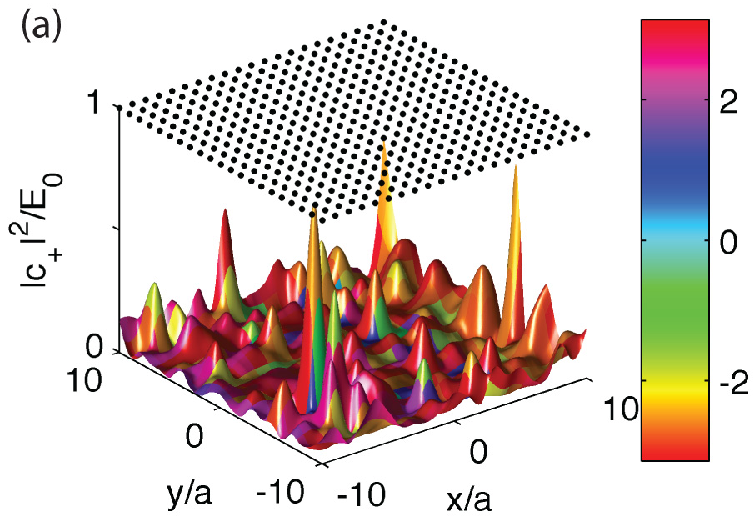}
  \includegraphics[width=0.9\columnwidth]{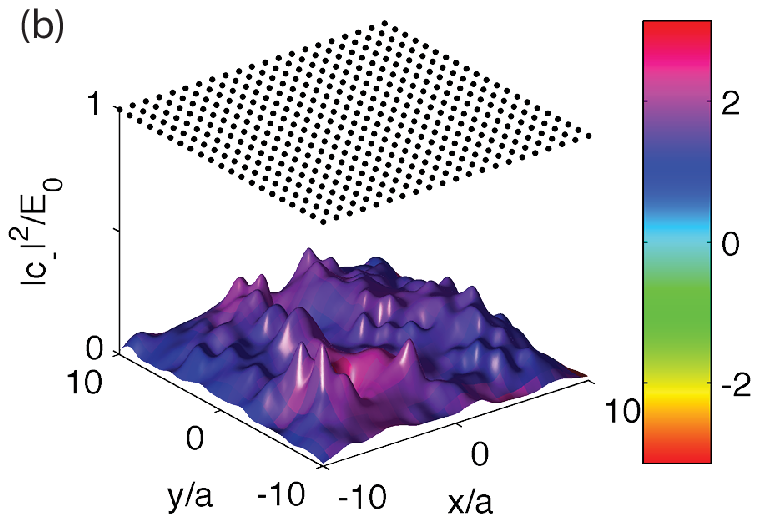}
  \caption{Response of inhomogeneously broadened square lattice to an
    incident plane wave resonant on the homogeneously broadened uniform magnetic
    mode $\colvec{v}_{\rmm}$.
    (a) The electric dipole excitations $|c_{+,l}|^2$ and (b) the
    magnetic dipole excitations $|c_{-,l}|^2$ of the response.
    The phases of the electric ($c_{+,l}$) and magnetic ($c_{-,l}$)
    dipole excitations are indicated by the color of the surfaces in
    (a) and (b), respectively.
    The black dots indicate the positions of the ASRs in the array.
    The individual ASR resonance frequencies are shifted by a
    independent identically distributed Gaussian random variables with
    standard deviation $\sigma = 0.8\delta\omega$.
    All other parameters are as in
    Fig.~\ref{fig:response-nonBroadened}.
  }
  \label{fig:response-Broadened}
\end{figure}

The introduction of inhomogeneous broadening alters the collective
interactions and can destroy the characteristics of the metamaterial
response that produces the transmission resonance.
We model the inhomogeneous broadening by shifting the central
resonance frequency of each ASR $l$ by independent identically
distributed Gaussian random variables $X_l$ with zero mean and
standard deviation $\sigma$.
With the asymmetry characterized by $\delta\omega$, the right and left
circular arcs in ASR $l$ possess resonance frequencies $\omega_{2l-1}
= \omega_0 + X_l +\delta\omega$ and $\omega_{2l}=\omega_0 + X_l -
\delta\omega$, respectively.
The deleterious effects of this broadening on the response are
illustrated in Fig.~\ref{fig:response-Broadened}, which shows a
much less uniform magnetic response in addition to localized electric dipole
excitations.
This non-uniformity inhibits the coherent reflection and
transmission through the meta-material array.

\begin{figure}
  \centering
  \includegraphics{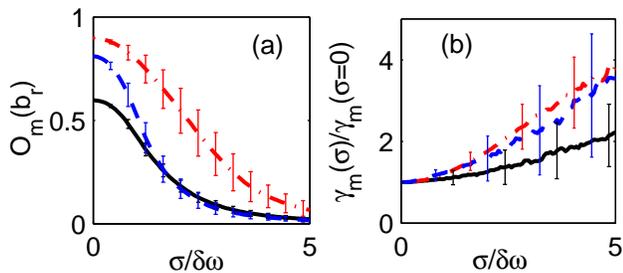}
  \caption{(a) The overlap of the uniform magnetic mode with the
    excitation driven by an incident EM plane wave and (b) the decay
    rate $\gamma_{\rmm}(\sigma)$ of the uniform magnetic mode
    $\colvec{v}_{\rmm}(\sigma)$ as a function of  inhomogeneous broadening
    $\sigma$.
    These quantities are plotted for $\delta\omega = 0.3\Gamma$ and
    ohmic losses $\Gamma_{\mathrm{O}} = 0.14\Gamma$ (solid black
    line), $\delta\omega = 0.3 \Gamma$ and $\Gamma_{\mathrm{O}} = 0$
    (dashed blue line), and $\delta\omega = 0.1 \Gamma$ and
    $\Gamma_{\mathrm{O} } = 0$ (dot dashed-red line).
    The lines indicate the average over $240$ sample realizations, and
    the error bars indicate the standard deviations.
    The incident wave has electric field polarization $\unitvec{e}_y$
    aligned with the ASR electric dipoles and is resonant on the mode
    $\colvec{v}_{\rmm}(\sigma=0)$ of the homogeneously broadened array.
    The decay rates of the uniform magnetic modes for a homogeneously
    broadened
    array $\gamma_{\rmm}(\sigma=0)$ are $0.308\Gamma$ (black line),
    $0.145\Gamma$ (blue line) and $0.0371\Gamma$ (red line).
    This shows that excitation of the magnetic mode, and hence the
    transmission resonance, vanishes as the inhomogeneous broadening
    becomes comparable to $\delta\omega$.
}
  \label{fig:modeOverlap}
\end{figure}

We quantify the effects of inhomogeneous broadening by examining the
overlap of the metamaterial steady state response $\colvec{b}_{\mathrm{r}}$ with
the magnetic mode $\colvec{v}_\rmm(\sigma=0)$ of a homogeneously
broadened array (corresponding to the case
in which all ASR meta-molecules are identical).
If the excitation is purely in the mode $\colvec{v}_\rmm(0)$, then the
overlap $O_\rmm(\colvec{b}_{\mathrm{r}})$ [Eq.~\eqref{eq:overlapDef}] is unity.
Figure~\ref{fig:modeOverlap}(a) shows the overlap
$O_\rmm(\colvec{b}_{\mathrm{r}})$ of
the response to an incident field resonant on the homogeneously broadened mode
$\colvec{v}_{\rmm}(\sigma=0)$ averaged over  $240$ realizations.
The solid black line was calculated for the same parameters as in
Figs.~\ref{fig:uniformElectricMode} through
\ref{fig:response-Broadened} with varying degrees of broadening.
The blue dashed curve shows the corresponding overlap in the absence
of ohmic losses, while for the red dashed-dot curve, the asymmetry
parameter was reduced to $\delta\omega=0.1$ and $\Gamma_O=0$.
In all cases, as the broadening standard deviation $\sigma$ becomes
comparable to $\delta\omega$, the ability to excite the uniform
magnetic mode drastically decreases.
The other modes that are excited either contain electric dipole
components or are not phase matched.
This either results in scattering of the field or in absorption of the
field due to ohmic losses.
As a result, the coherent collective response responsible for the
transmission resonance observed in by Fedotov \textit{et al}
\cite{FedotovEtAlPRL2010} becomes unobservable when the inhomogeneous broadening is larger than $\delta\omega$.
These collective effects do persist, however, for $\sigma$ roughly half
$\delta\omega$.
We show the effect of inhomogeneous broadening in the decay rate
$\gamma_{\rmm}(\sigma)$ of
the magnetic mode itself in Fig.~\ref{fig:modeOverlap}(b) as a function
of broadening.
Randomization of the ASR resonance frequencies apparently has little
effect on the collective linewidth of the magnetic mode for $\sigma <
\delta\omega$.
However, for larger degrees of broadening, the decay rate of this mode
can be increased several times over and the resonance linewidth narrowing,
that results from the cooperative response of the metamaterial array, disappears.
Furthermore, the large standard deviations of $\gamma_\rmm(\sigma)$
indicate that the width of the uniform magnetic mode is highly
sensitive to the particular realization of ASR resonance frequencies.
A larger decay rate renders the magnetic mode more difficult to excite
since any excitation of this mode is more quickly radiated away.
In order to for an array of ASRs to exhibit a transmission resonance,
a large fraction of the excitation created by the driving field must
be in the uniform magnetic mode distributed over the array.
In low loss metamaterials, this can be achieved in conjunction with a
higher quality for that resonance for larger arrays and smaller values
of $\delta\omega$. \cite{JenkinsLineWidthArxiv}
However, as Fig.~\ref{fig:modeOverlap} indicates, reduction in the
asymmetry to achieve this quality factor enhancement correspondingly
reduces the tolerance for inhomogeneous broadening in the resonance
frequency.

The observation that the cooperative metamaterial response to EM fields
can be suppressed in the presence of sufficiently strong inhomogeneous broadening
could potentially also be exploited in design of metamaterial samples that would benefit from
well-defined homogeneous properties for electric susceptibility and magnetic permeability,
such as diffraction-free lenses due to negative refractive index.\cite{SmithEtAlPRL2000,ShelbySci2001,SmithEtAlSCI2004}
One could prepare a controlled amount of inhomogeneous broadening for the metamaterial
sample in order to generate an EM response that more closely mimics standard continuous medium electrodynamics with suppressed contribution from recurrent scattering events.

\section{Conclusion}
\label{sec:conclusion}

In conclusion, we have analyzed the collective modes of a finite-sized
2D metamaterial array of ASR resonators and how they are influenced by
an inhomogeneous broadening of the resonance frequencies of the
individual resonators.
The study was motivated by recent experimental observations of
transmission resonance linewidth narrowing as a function of the size
of the system.\cite{FedotovEtAlPRL2010}
This effect can be understood by analyzing the resonance linewidths of
collective modes of the system that undergo dramatic narrowing due to
strong EM field mediated interactions between the resonators.
As demonstrated by previous comparisons between the numerical
simulation results and the experimental
observations,\cite{JenkinsLineWidthArxiv} the response can be analyzed
by a simplified model in which each meta-atom is treated as a discrete
element supporting a single mode of current oscillation possessing
electric and magnetic dipole moments.
Collective interactions between the meta-atoms are mediated by the
scattered EM fields.
The excellent agreement between the theory and the experiment can be
understood by a relatively weak higher-order multiple radiation of
individual ASR metamolecules.\cite{PapasimakisComm}

We examined in detail how inhomogeneous broadening of resonator
resonance frequencies impairs the coherent collective phenomena that
are expected to find important applications in metamaterial systems.\cite{ZheludevEtAlNatPhot2008,KAO10,SentenacPRL2008,FedotovEtAlPRL2007}
While the transmission resonance experimentally observed in
Ref.~\onlinecite{FedotovEtAlPRL2010} persists for inhomogeneous
broadening that is a fraction of the ASR asymmetry parameter
$\delta\omega$, the cooperative response vanishes when the broadening
begins to exceed that parameter.
Production of high quality resonances with low loss materials requires
the reduction of $\delta\omega$.\cite{JenkinsLineWidthArxiv}
Figure~\ref{fig:modeOverlap} indicates that, in order to produce such
high quality resonances, the uniformity in the production of
meta-molecules will need to become correspondingly small.

\begin{acknowledgements}
  We acknowledge discussions with
  N.\ Papasimakis, V.\ Fedotov, and N.\ Zheludev and financial support from the EPSRC and the Leverhulme Trust.
\end{acknowledgements}

%

\end{document}